\documentclass[12pt]{article}
\usepackage{amssymb}

\newcommand{\be}{\begin{equation}}
\newcommand{\ee}{\end{equation}}
\newcommand{\bn}{\begin{eqnarray}}
\newcommand{\en}{\end{eqnarray}}

\newcommand{\nn}{\nonumber}

\def\bea{\begin{eqnarray}}
\def\eea{\end{eqnarray}}

\begin{document}

\author{D. Dalmazi$^{b}$\thanks{%
E-mail: dalmazi@feg.unesp.br} and A. de Souza Dutra$^{a,b}$\thanks{%
E-mail: dutra@feg.unesp.br} \\
$^{a}$Abdus Salam ICTP, Strada Costiera 11, 34014 Trieste Italy\\
$^{b}$UNESP - Campus de Guaratinguet\'{a} - DFQ\thanks{%
Permanent Institution}\\
Av. Dr. Ariberto Pereira da Cunha, 333\\
12516-410 Guaratinguet\'{a} SP Brasil}
\title{{\LARGE Restrictions over two-dimensional gauge models with Thirring-like
interaction.}}
\maketitle

\begin{abstract}
Some years ago, it was shown how fermion self-interacting terms of the
Thirring type impact the usual structure of massless two-dimensional gauge
theories \cite{st}. In that work only the cases of pure vector and pure
chiral gauge couplings have been considered and the corresponding Thirring
term was also pure vector and pure chiral respectively, such that the vector
(or chiral) Schwinger model should not lose its chirality structure due to
the addition of the quartic interaction term. Here we extend this analysis
to a generalized vector and axial coupling both for the gauge interaction
and  the quartic fermionic interactions. The idea is to perform quantization
without losing the original structure of the gauge coupling. In order to do
that we make use of an arbitrariness in the definition of the Thirring-like
interaction.

\end{abstract}

\newpage

The physics of the two-dimensional quantum field theories has a long and
beautiful history. Particularly, one of the most important and beautiful
features of this restricted number of dimensions is that it allows the exact
solvability of some important fermionic models, which is intrinsically
connected with the idea of bosonization. The bosonization program leads to a
deeper understanding of the internal structure of the models. People have
been trying to extend it for higher dimensions for a long time. One of the
first models to be exactly solved in the history of the quantum field theory
is the Thirring model \cite{thirring}, and a little time after that
Schwinger presented to the world what is now a landmark for the bosonization
technique, the so called Schwinger model \cite{schwinger}, whose solution
through operator approach was carried out by Lowenstein and Swieca in
another beautiful and classical paper \cite{swieca}. Its path integral
solution was obtained in \cite{gamboa} by following an approach introduced
by Fujikawa \cite{fujikawa}. After several works analyzing a number of
properties and applications of these and other exact two-dimensional
systems, Jackiw and Rajaraman \cite{jackiw} published a seminal paper in
this subject where they have presented the first consistent model with an
anomalously broken gauge symmetry, the chiral Schwinger model (CSM). In a
non-trivial extension, some authors developed these ideas for the case of
the arbitrary left and right couplings which may be called a generalized
Schwinger model (GSM) \cite{chanowitz,boyanovsky,ijmp95}.

Some years ago \cite{st}, some of the above ingredients were combined
together in order to explore the effect of the quartic fermionic
self-interaction of the Thirring type on, among other physical properties,
the dynamically generated photon mass. It has been shown in that work that,
for instance in the case of the Schwinger model, the dynamically generated
photon mass is screened due to the self-interacting Thirring term as follows:

\begin{equation}
m^2=\frac{e^2}{\pi+g^2},
\end{equation}

\noindent We can also speak of a kind of duality, in the sense that a model
with strong Thirring coupling $g$ can be traded in another one with small
electromagnetic coupling $e$ \cite{st}.

In this work we initially extend the analysis of the impact of fermionic
self-interactions on the structure of the gauge theories in $1+1$ dimensions
and, after demonstrating that the special case of CSM can not be taken as a
limit of the general model, we present the difficulties in performing this
task without losing the chirality of the original CSM model.

\section{Generalized Schwinger model with Thirring coupling}

Our starting point is a generalized Schwinger model, below $P_{\pm}= (1\pm
\gamma_5)/2$, :

\begin{equation}
\mathcal{L}=\bar{\Psi}(i\partial \!\!\!/+e_{R}\,\gamma ^{\mu }\,P_{+} A_{\mu
}+e_{L}\,\gamma ^{\mu }\,P_{-}A_{\mu })\Psi - \,\frac{1}{4}F^{\mu \nu
}F_{\mu \nu }.  \label{gsm}
\end{equation}

\noindent Next we would like to add a general quartic fermionic interaction
such that it is both chiral and gauge invariant while keeping the explicit
Poincare covariance. Before we introduce our model it is important for our
purposes to make use of a two dimensional spinor identity proved in {\cite
{mig}}, namely,

\begin{equation}
\left(\bar{\Psi} M \Psi \right)^2 = \det M \left(\bar{\Psi} \Psi \right)^2
\qquad .  \label{id1}
\end{equation}

\noindent Where $M$ may be any $2\times 2$ nonsingular matrix. Now, it
follows from $\det P_{\pm} = 0 $ and (\ref{id1}) that $\left(\bar{\Psi}
\gamma_{\mu} P_{\pm} \Psi \right)^2 = 0 $. Consequently, by making use of
the representations $\bar{\Psi} \gamma_{\mu} \Psi = \bar{\Psi}
\gamma_{\mu}(P_+ + P_-) \Psi $ and $\bar{\Psi} \gamma_{\mu}\gamma_5 \Psi =
\bar{\Psi} \gamma_{\mu}(P_+ - P_-) \Psi $ we can derive

\begin{equation}
a \left( \bar{\Psi} \gamma_{\mu} \Psi \right)^2 + b \left( \bar{\Psi}
\gamma_{\mu}\gamma_5 \Psi \right)^2 + c \left( \bar{\Psi} \gamma_{\mu} \Psi
\right)\left( \bar{\Psi} \gamma^{\mu}\gamma_5 \Psi \right) = \frac{\left( a
- b \right)}{e_R e_L} \left\lbrack \bar{\Psi} \gamma_{\mu}\left( e_R P_+ +
e_L P_- \right)\Psi \right\rbrack^2.  \label{id2}
\end{equation}

\noindent Where the constants $a,b,c$ are arbitrary and we have assumed $%
e_R,e_L \ne 0$. Note that the right handed side of (\ref{id2}) is
independent of $c$ which stems from the fact that the term $j^{\mu}j_{\mu}^5$
which multiplies $c$ vanishes identically by virtue of (\ref{id1}) and the
afore mentioned representations. The vanishing of both sides of (\ref{id2})
for $a=b$ can be easily checked from (\ref{id1}) and $\det \gamma_5 =-1$.

Therefore, with the exception of the chiral Schwinger models $e_R=0$ or $%
e_L=0$, the identity (\ref{id2}) allows us to add a rather general quartic
fermionic term to the generalized Schwinger model in a such way that the
chiral structure of the gauge coupling is maintained. That is,

\begin{eqnarray}
\mathcal{L}&=& \bar{\Psi}(i\partial \!\!\!/+e_{R}\,\gamma ^{\mu
}\,P_{+}\,A_{\mu }+e_{L}\,\gamma ^{\mu }\, P_{-}\,A_{\mu })\Psi \,-\,\frac{%
g^{2}}{2}\left( \bar{\Psi}\gamma _{\mu }\left( e_{R} P_{+}+
e_{L}P_{-}\right) \Psi \right) ^{2}  \nonumber \\
&-& \,\frac{1}{4}F^{\mu \nu }F_{\mu \nu } \, + \, j^{\mu} \bar{\Psi}\gamma
_{\mu }\left( e_{R} P_{+}+ e_{L}P_{-}\right) \Psi ,  \label{gsmth}
\end{eqnarray}

\noindent where we have introduced a source term for future use. In order to
deal with this model, one usually reduces the order of the quartic
interaction by means of an auxiliary vector field $B_{\mu}$ which leads us
to a physically equivalent effective theory

\begin{eqnarray}
\mathcal{L}_{eff}&=& \bar{\Psi}\left(i\partial
\!\!\!/+e_{R}\,\gamma ^{\mu }\,P_{+}\,A_{\mu }+e_{L}\,\gamma ^{\mu
}\,P_{-}\,A_{\mu }\right)\Psi - g B_{\mu}
\bar{\Psi}\left(e_{R}\,\gamma ^{\mu }\,P_{+} + e_{L}\gamma ^{\mu
}\,
P_{-}\right)\Psi + \frac 12 B_{\mu}B^{\mu}  \nonumber \\
&-& \,\frac{1}{4}F^{\mu \nu }F_{\mu \nu } \, + \, j^{\mu} \bar{\Psi}\gamma
_{\mu }\left( e_{R} P_{+}+ e_{L}P_{-}\right) \Psi .  \label{gsmth1}
\end{eqnarray}

\noindent By integrating over $B_{\mu}$ in the path integral or using its
equations of motion we recover (\ref{gsmth}). Now it is clear that the
fermionic fields interact with the gauge field, the auxiliary field and the
source through the combination:

\begin{equation}
C_{\mu }\equiv A_{\mu } + j_{\mu} - g\,B_{\mu }\,, \label{cb}
\end{equation}

\noindent in terms of which, the Lagrangian density can be rewritten as
\begin{equation}
\mathcal{L}_{eff}=\bar{\Psi}(i\partial \!\!\!/+ e_{R}\, \gamma ^{\mu
}\,P_{+}\,C_{\mu }+e_{L}\,\gamma ^{\mu }\,P_{-}\,C_{\mu })\Psi +\frac{1}{2
g^2}\left(A_{\mu } + j_{\mu} - C_{\mu}\right)^2 - \,\frac{1}{4}F^{\mu \nu
}F_{\mu \nu } \;.
\end{equation}

\noindent This last expression can be bosonized through standard techniques
as in \cite{ijmp95} with the following steps: First of all, one introduces a
general vector decomposition $C_{\mu} = \epsilon_{\mu\nu}\partial^{\nu}\chi
+ \partial_{\mu}\eta $. Next, one decouples the fermionic fields through a
chiral and a gauge transformation:

\begin{equation}
\Psi = \exp \left\lbrack i \, e_V (\eta + \chi\gamma_5) + i \, e_A
(\eta\gamma_5 - \chi)\right\rbrack \zeta
\end{equation}

\noindent where

\begin{equation}
e_V = e_R+e_L \quad ; \quad e_A = e_R-e_L  \label{eva}
\end{equation}

\noindent The chiral transformation has a nontrivial Jacobian
\cite{fujikawa}. The effective theory including the Jacobian
becomes

\begin{eqnarray}
&&\mathcal{L}_{eff}=\bar{\zeta}(i\partial \!\!\!/)\zeta
+\frac{1}{2}\,\chi \Box ^{2}\chi +\frac{1}{2\,\pi }\left(
\frac{e_{A}^{2}}{4}-m^{2}\right) \eta \Box
\eta +   \\
&&\frac{1}{2\,\pi }\left( \frac{e_{V}^{2}}{4}+m^{2}\right) \chi \Box \chi -%
\frac{e_{V}\,e_{A}}{4\,\pi }\eta \Box \chi + \nn \\
&&+\frac{1}{2g^{2}}\left( A_{\mu }+j_{\mu }-\epsilon _{\mu \nu
}\partial ^{\nu }\chi -\partial _{\mu }\eta \right)
^{2}-\,\frac{1}{4}F^{\mu \nu }F_{\mu \nu }\nn\; ,
\end{eqnarray}

\noindent where $m$ is an arbitrary parameter with mass dimension.
The terms which depend on $\chi$ and $\eta$ are combined back into
the field $C_{\mu}$. Some of them can be written as a functional
integral over a scalar field $\phi$. After all those steps we
obtain the intermediate effective Lagrangian density,

\begin{eqnarray}
&&\mathcal{L}_{eff} = \bar{\zeta}(i\partial \!\!\!/)\zeta
+\frac{1}{2}\left(
\partial _{\mu }\phi \right) ^{2}-\frac{1}{2\sqrt{\pi }}\left\lbrack e_A\,
g^{\mu \nu }- e_V \,\epsilon^{\mu \nu }\right\rbrack \partial
_{\nu }\phi
\,C_{\mu }+  \nonumber \\
&&\frac{m^{2}}{2\,\pi }\,C_{\mu }^{2}+ \frac{1}{2 g^2}\left(A_{\mu } +
j_{\mu} - C_{\mu}\right)^2-\,\frac{1}{4}F^{\mu \nu }F_{\mu \nu }\; ,
\end{eqnarray}

\noindent  At this point we can change back the variables from the
field $C_{\mu}$ to $B_{\mu}$ according to (\ref{cb}). Integrating
out the auxiliary field $B_{\mu}$ and the free Dirac field $\zeta
$ we obtain a full bosonized Lagrangian density generalizing the
work of \cite{st}:

\begin{eqnarray}
\mathcal{L}_{\mathrm{bosonic}} &=& \frac{1}{2}\left[ 1+\frac{e_R \, e_L\, g^2%
}{\pi + m^2 g^2} \right] \left( \partial _{\mu }\phi \right) ^{2} -\frac{%
\sqrt{\pi}\left\lbrack e_A \,g^{\mu \nu }- e_V \,\epsilon ^{\mu
\nu }\right\rbrack \partial _{\nu }\phi}{2\left(\pi + m^2
g^2\right)} \, \left( A_{\mu } + j_{\mu} \right)
\nonumber \\
&+& \frac{m^{2}}{2\,\left(\pi + m^2 g^2\right) } \, \left( A_{\mu } + j_{\mu}\right)^2 -\,\frac{1%
}{4}F^{\mu \nu }F_{\mu \nu }\, \, .  \label{lbj}
\end{eqnarray}

\noindent Note that for $g=0$ the expression (\ref{lbj}) at $j_{\mu}=0$
reproduces the result of \cite{ijmp95} for the GSM.

Now we analyze the particle content of the bosonized theory. Integrating
over the scalar field in the absence of external sources we obtain an
effective nonlocal action for the photon:

\begin{equation}
\mathcal{L}_{eff}(A) = -\frac 14 F^{\mu\nu}\left\lbrack 1 + \frac{\alpha \,
(e_V^2 + e_A^2)}{4 \,\Box}\right\rbrack F_{\mu\nu} + \left( \frac {m^2}{K} -
\frac{\alpha \, e_A^2}4\right) \frac{A_{\mu}A^{\mu}}2 + \frac{\alpha\, e_V
\, e_A}4 \epsilon^{\mu\nu}\partial_{\mu}A_{\nu}\frac 1{\Box}
\partial_{\beta}A^{\beta} ,  \label{lefa}
\end{equation}

\noindent where we have the constants:

\begin{equation}
\quad K = \pi + m^2 g^2 \quad ; \quad \alpha = \frac {\pi}{K(K+e_R \, e_L
g^2 )} .  \label{k}
\end{equation}

\noindent In the vector case $e_R=e_L\equiv e $ we can recover the gauge
symmetry by fixing conveniently the arbitrary parameter $m^2 =0$  which
leads, redefining $g\to g/e$, to a nonlocal gauge theory with a massive
photon $m_{ph}^2 = e^2/(\pi + g^2)$ which reproduces the result of \cite{st}
and the result of the Schwinger model for $g=0$. In the general case the
gauge symmetry is broken, so we can compute the propagator without gauge
fixing. We first notice that $\mathcal{L}_{eff}(A)=
A_{\mu}G^{\mu\nu}A_{\nu}/2$ where the kinetic operator is of the form $%
G_{\mu\nu}=a \, g_{\mu\nu} + b\, \theta_{\mu\nu} + c\, V_{\mu\nu}$ with the
definitions in momentum space $\theta_{\mu\nu}= g_{\mu\nu}
-k_{\mu}k_{\nu}/k^2$, $V_{\mu\nu} = T_{\mu\nu} + T_{\nu\mu}$, with $%
T_{\mu\nu}=\epsilon_{\alpha\mu}k_{\nu}k^{\alpha}$. It is easy to check that
the propagator is given by:

\begin{equation}
\left\langle A_{\mu}(k) A_{\nu}(-k) \right\rangle \, = \, G^{-1}_{\mu\nu} =
\frac 1D \left\lbrack (a+b)\, g_{\mu\nu} - b \, \theta_{\mu\nu} - c \,
V_{\mu\nu} \right\rbrack ,  \label{propa}
\end{equation}

\noindent where

\begin{equation}
a = \frac {m^2}{K} - \frac{\alpha \, e_A^2}4 \quad ; \quad b = \frac{%
\alpha\, (e_V^2 + e_A^2)}4 - k^2 \quad ; \quad c = \frac{\alpha \, e_V \, e_A%
}{4 k^2} \qquad ,  \label{abc}
\end{equation}

\begin{equation}
D = (a+b)a + k^4 c^2 = k^2 \left( \frac{\alpha \, e_A^2}4 +\frac
{m^2}{K}\right) - \frac {m^2}{K}\left\lbrack \alpha\, e_R\, e_L + \frac
{m^2}{K}\right\rbrack .  \label{D}
\end{equation}

\noindent The last term in the propagator (\ref{propa}) contains  a pole at $%
k^2=0$. However, if we saturate the propagator with  conserved currents and
calculate the residue at this pole we have a vanishing result,  i.e., $%
\lim_{k^2\to 0} k^2 J_{\mu}\left\langle A^{\mu}(k) A^{\nu}(-k) \right\rangle
J_{\nu}^* =0 $ . Therefore we do not have a  propagating massless particle
as one might think. On the other  hand if we repeat this calculation at the
pole coming from the condition $D=0$,  we have a positive residue indicating
a physical massive photon $k^2=m_{ph}^2$  whose mass depends on the
arbitrary regularization parameter as  follows:

\begin{equation}
m_{ph}^2 = \frac{m^2 \left\lbrack m^2 + \left(e_V^2 -
e_A^2\right)/4\right\rbrack}{m^2 g^2 \left\lbrack m^2 + \left(e_V^2 -
e_A^2\right)/4\right\rbrack + \pi \left( m^2 - e_A^2/4\right) }  \label{mph}
\end{equation}

\noindent The result (\ref{mph}) is both in agreement with the chiral case $%
(e_V=e_A)$ treated in \cite{st} and the GSM without Thirring interaction $%
(g^2=0)$ studied in \cite{naon}. As in those cases, the arbitrary parameter $%
m^2$ must satisfy a specific bound in order to avoid tachyons.

Finally, we return to the issues of bosonization and the chiral limit. Since
the quadratic term in the sources in (\ref{lbj}) can only give rise to
contact terms in the correlation functions of the $U(1)$ current, by
comparing (\ref{lbj}) with (\ref{gsmth}) we read off the bosonization rule:

\begin{equation}
\bar{\Psi}\gamma ^{\mu }\left( e_{R} P_{+}+ e_{L}P_{-}\right) \Psi = \bar{%
\Psi}\gamma ^{\mu }\left( e_{V} + e_{A}\gamma_5 \right) \Psi = \frac{m^{2}}{%
\left(\pi + m^2 g^2\right) } \,A^{\mu } -\frac{\sqrt{\pi}\left\lbrack e_A
\,g^{\mu \nu }- e_V\,\epsilon ^{\mu \nu }\right\rbrack \partial _{\nu }\phi}{%
2\left(\pi + m^2 g^2\right)} .  \label{jb}
\end{equation}

\noindent As in the chiral case studied in \cite{st}, we remark
that the bosonization rule for the above generalized current is
ambiguous due to the arbitrary, regularization dependent,
parameter $m^2$. As commented in \cite {st}, we notice that the
identity $\left(\bar{\Psi}\gamma ^{\mu } P_{\pm} \Psi \right)^2
=0$ will not be respected by the bosonic map (\ref{jb}) in the
chiral cases $e_R=0$ or $e_L=0$. This is a quantum effect, since
it is caused by the gauge field term on the right handed side of
(\ref{jb}) which appears due to the regularization procedure. In
fact, had we started with the CSM plus a chiral Thirring
interaction:

\begin{equation}
\mathcal{L}=\bar{\Psi}\left\lbrack i\partial \!\!\!/+ e\,P_{\pm
}\,\gamma ^{\mu }(A_{\mu }+ j_{\mu})\right\rbrack\Psi
-\,\frac{1}{4}F^{\mu \nu }F_{\mu \nu }\,-\,\frac{g^{2}}{2}\left(
\bar{\Psi}\gamma _{\mu }\left( P_{\pm }\right) \Psi \right) ^{2},
\label{l2}
\end{equation}

\noindent since the last term vanishes identically, see comment
after (3), we would end up with another bosonized Lagrangian
density different from (\ref{lbj}):
\begin{equation}
\mathcal{L}_{\mathrm{bosonic}}(a) = \frac{1}{2}\left(
\partial _{\mu }\varphi \right) ^{2}-\frac{e}{\sqrt{\pi }}\left(
\,g^{\mu \nu }\,\pm \,\epsilon ^{\mu \nu
}\right) \partial _{\nu }\varphi \, (A_{\mu }+j_{\mu}) + \frac{a \, e^{2}}{2\,\pi }%
\, (A_{\mu }+j_{\mu})^{2} - \frac 14 F_{\mu\nu}^2 \, ,
\label{lbj2}
\end{equation}

\noindent where $a$ is the Jackiw-Rajaraman \cite{jackiw}
parameter which represents regularization ambiguities. Analogous
to the derivation of (\ref{jb}) we derive from (\ref{l2}) and
(\ref{lbj2}) a new bosonization rule which is also ambiguous but
differs from (\ref{jb}), namely,

\be e \, \bar{\Psi}\gamma ^{\mu } P_{\pm} \Psi (a) = \frac{a \,
e^{2}}{\pi } \, A^{\mu } -\frac{e}{\sqrt{\pi }}\left( \,g^{\mu \nu
}\,\pm \,\epsilon ^{\mu \nu }\right)
\partial _{\nu }\varphi \quad . \label{jb2} \ee

There is of course no trace of the Thirring interaction in
(\ref{lbj2}) and (\ref{jb2}). On the other hand, in the chiral
limits $(e_R,e_L) \to (0,e)$ or $(e_R,e_L) \to (e,0)$, the
Lagrangian density (\ref{gsmth})  coincides with (\ref{l2}) after
the trivial replacement $g \to g \, e_R $ in  (\ref{l2}). However,
differently from  (\ref{lbj2}) and (\ref{jb2}) we have some
dependance on the Thirring coupling constant in (\ref{lbj}) and
(\ref{jb}) in the chiral limits. Even if we redefine the arbitrary
parameter $m^2$ such that $m^2/(\pi + m^2 g^2) \to a\, e^2 $ we
still have a mismatch between formulas (\ref{lbj}) and (22)
and also (\ref{jb}) and (23). The difference might be
interpreted as a finite charge renormalization. Furthermore, we
should identify $\phi = \varphi$ for $e_L=0$ while for $e_R=0$ we
have $\phi = - \varphi$. We can only make the couple of equations
(\ref{lbj}),(\ref{jb})  coincide with (\ref{lbj2}),(\ref{jb2}) in
the chiral limit either by taking $g^2 \to 0$ or selecting out the
parameters $\, m^2 = 0 = a\, $. As a consequence we should
conclude that there is apparently no way of introducing the
Thirring-like interaction for the CSM without losing its chiral
characteristic. In fact, it can be shown that, for a large class
of non-chiral Thirring-like interactions, the chirality of the
original fermions is lost as it should be expected. Finally, a
quite interesting and intriguing conclusion from above is that one
can introduce the Thirring-like interaction consistently,
preserving the original mixture of right-left modes, except for
the CSM models.

\section{Acknowledgments}

The authors are grateful to CNPq for partial financial support. This work
has been finished during a visit (ASD) within the Associate Scheme of the
Abdus Salam ICTP.

\end{document}